# THz emission from multiply ionized laser plasma

Lucie Jurkovičová[1,2,*], David Štok[1,2], Caroline Juliano[1], Matyáš Staněk[1,2], Jaroslav Nejdl[1,2], and Ondřej Hort[1,*]
[1]ELI Beamlines Facility, The Extreme Light Infrastructure ERIC, Za Radnicí 835, 25241 Dolní Břežany
[2]Czech Technical University in Prague, FNSPE, Břehová 7, 115 19 Prague 1, Czech Republic.
*Address correspondence to: Lucie.Jurkovicova@eli-laser.eu and Ondrej.Hort@eli-laser.eu

**Abstract**
Studies employing nonlinear interactions of ultrashort THz pulses with matter are becoming a promising scientific research field enabling access to ultrafast dynamics unattainable by other methods. To enter this regime, THz pulses with energy ranging from hundreds of μJ to the mJ level are necessary. However, techniques that provide pulses with such energy levels are still not widely established. Upscaling methods of laser-solid interaction is limited by the damage threshold of materials, while the mechanism of THz generation from high-intensity laser-gas interactions is not fully understood yet. Here, we establish the photocurrent model of laser-driven plasma THz generation in the unexplored high-intensity regime by accounting for high-ionization states of the target gas. Our model shows excellent agreement with experimental observations, providing a clear explanation of phenomena in both spectral and temporal domains. Moreover, in such a regime it explains a newly observed significant increase in conversion efficiency from laser to THz. In the experiments, we achieved a generation of 0.2 mJ THz pulses, driven by a Ti:sapphire laser with a conversion efficiency exceeding 1 %.

# INTRODUCTION

Ultrafast nonlinear terahertz (THz) studies [1, 2, 3] such as THz induced nonlinear processes in crystals [4, 5, 6], THz induced superconductivity [7, 8], THz assisted high-order harmonic generation [9, 10, 11, 12], attosecond pulse generation [13, 14] or enhancement in particle acceleration [15, 16, 17] represent some of the wealth of promising applications of high-peak power THz pulses. However, powerful THz sources enabling nonlinear effects are still not widely available.

Extraordinary progress made in the optical rectification techniques reaching pulse energies close to 1 mJ and 10 % of conversion efficiency [18, 19, 20] is encouraging, however, due to long pulse durations in the order of picoseconds and bandwidth limited to $< 10$ THz [18], this method has not been suitable to reach very high peak-powers so far.

Another possibility of THz generation is based on two-color laser plasma, which offers unique properties of the THz radiation such as femtosecond pulse duration [21] and corresponding bandwidth spanning from few tens up to hundreds of THz [22, 23]. However, this comes at the cost of lower conversion efficiency, reaching THz pulse energies in the range from tens of μJ for near-IR (NIR) drivers up to few hundreds of μJ for mid-IR laser systems [24, 25, 26, 27, 28, 29].

Recent advances in laser technology [30, 31, 32, 33] have paved the way for high-energy [27] and high-repetition rate THz sources [34], making the two-color laser plasma generation a promising technique. Unlike in optical rectification, the damage threshold of nonlinear materials does not limit this approach, making it an ideal platform to upscale THz generation.

Many techniques have been presented to improve the generation in two-color laser plasma: using double pulses [35], phase modulations in two-color driving pulse [36], spatio-temporal coupling of the laser pulse [37, 38], abruptly focusing beams [39], or controlling polarizations of the two-color field [40]. Even though the energy upscaling of THz generation was already proposed [41, 22], all these techniques are exploring properties of the THz generation in the regime of moderate intensities ($\approx 10^{14}\,\mathrm{W/cm^2}$) of the driver where the generation mechanism is well explained by the single-ionization photocurrent model [42]. In this regime, it was shown that the THz spectrum bandwidth broadens with shortening of the driving pulse duration and that the THz pulse duration follows approximately the duration of the driving laser [43]. However, in the high-intensity regime using femtosecond multi-mJ driving laser pulses, the understanding of spectral and temporal properties of the generated THz radiation has remained limited.

Here, we shed light on the generation of THz radiation in a two-color laser plasma appearing in the high intensity of the driving pulses reaching the underexplored region up to the $10^{17}\,\mathrm{W/cm^2}$. We present a THz source based on a two-color laser plasma driven by a 20 mJ Ti:sapphire laser system, yielding up to 220 µJ of THz pulse energy in krypton gas. This represents a conversion efficiency of 1.1 %, one of the highest reported for this type of driving laser [29, 37]. To understand the process, we established a new photocurrent model including high ionization states of target noble gases, going up to $Kr^{10+}$ and $Ne^{8+}$ ions. We show that our model fully corresponds to the experimental data and can explain observed complex features – position of the spectral peak and increase in conversion efficiency. Moreover, our model provides a clear and valuable insight into the temporal structure of the generated THz pulses. From the experimental observations supported by our model, we show that the spectral and energy scaling laws depend only on the ionization dynamics of the medium, providing important experimental insight into the otherwise elusive nature of ionization by intense laser pulses – a topic that extends far beyond the scope of THz generation.

# METHODS

## Generation of a bi-chromatic driving laser beam

To study the THz generation experimentally, we performed a series of measurements with pure noble gases at various pressures providing a well-defined medium for two-color laser plasma THz generation. We have used a Ti:sapphire laser system with central wavelength of 800 nm, delivering laser pulse energies up to 55 mJ, compressed to 40 fs at a repetition rate of 10 Hz. Fig. 1a shows the schematics of the THz generation setup.

First, a few percent of second harmonic (SH) of the NIR beam is generated in a 200 µm thick beta-barium borate (BBO) type 1 crystal, cut at a phase-matching angle of 29.2°. The thickness of the crystal is optimized to preserve the pulse duration of 40 fs of the driving NIR pulse. As the crystal is placed in the collimated beam and therefore in low intensity, the measured second harmonic conversion efficiency varied between 2 % and 5 % for the different laser energies used. The other factor decreasing the conversion efficiency was the optimization of the temporal compression of the driving beam to be the shortest pulse at the position of THz generation after 7.5 mm of glass (Calcite, WP and W1 – see Fig. 1a). This resulted in a lower NIR intensity in the crystal due to slightly chirped pulses leading to a lower conversion efficiency of the SH generation. A 2.4 mm thick calcite crystal, cut at an angle of 55° was used to compensate for the group delay between the NIR and SH pulses for a temporal overlap inside the generation medium and fine-tuned to set the phase delay between them to $\varphi = {}^{\pi}/_{2}$ in order to maximize the generation [44]. A thin true zero-order waveplate (45 μm thick quartz) working as a half-waveplate for 800 nm and full-waveplate for 400 nm is inserted into the beam to set the perpendicular polarization of the NIR to become colinear with the SH to enhance the generation efficiency [42]. The bi-chromatic beam enters a gas cell through a 5 mm thick fused silica window.

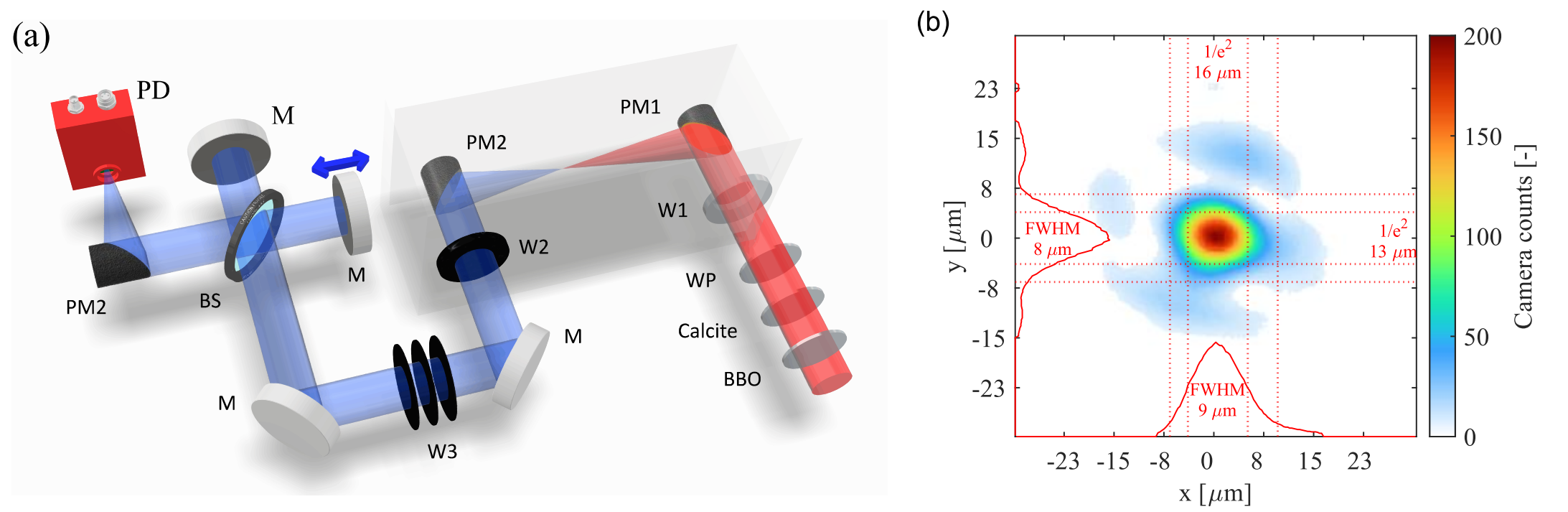

**Fig. 1 Experimental setup for THz generation in two-color laser plasma in noble gases. (a)** The collimated NIR laser pulse first drives a second harmonic generation in beta-barium borate (BBO) doubling crystal to generate the bi-chromatic laser field. A calcite crystal serves to adjust the group delay between the two colors and a true zero-order half-waveplate (WP) is used to rotate the second harmonic polarization to be colinear with the NIR. The bi-chromatic laser beam enters a gas cell by a fused silica window (W1) and is tightly focused by a silver parabolic mirror (PM1). The generated THz radiation is collimated by a gold parabolic mirror (PM2), passes through a HRFZ-Si window (W2) and optionally through the set of thin HRFZ-Si attenuation plates (W3) and is conducted by gold mirrors (M) to a Fourier-transform spectrometer using a pellicle beamsplitter (BS). A calibrated pyroelectric detector (PD) WiredSense MPY-RS with a diamond window is used to detect the THz signal. **(b)** Laser focal spot of the NIR beam measured under the vacuum inside the gas cell for $f_\# = 7$.

## Focal spot size measurement

The bi-chromatic beam is focused inside the gas cell by a silver off axis parabolic mirror with $f_\# = 5$. To understand the THz generation, it is essential to properly calibrate the intensity of the laser field reached in the focus. To achieve this, we removed the second collimating parabolic mirror and propagated the diverging beam after the focal spot out of the cell through another fused silica window. Next, we imaged the focal spot to the camera using an achromatic doublet lens with a focal length $f = 300$ mm and magnification factor of $M = 5.46$. By this method we obtained a focal spot size image for different energies of the laser and different iris sizes. Typical energy contained in the central lobe of the focal spot varied between 65 % and 90 % of the total laser pulse energy for the largest irises and the smallest irises respectively. Fig. 1b shows the focal spot inside the cell under the vacuum conditions measured with a slightly clipped driving beam by the iris to $f_\# = 7$.

## Detection of THz radiation

The generated THz beam is collimated by a gold parabolic mirror with $f_\# = 3$ and passes through a high-resistivity float-zone silicon (HRFZ-Si) window that serves also as a blocking filter for the residual NIR and SH. The THz beam passes through an optional set of thin HRFZ-Si plates to attenuate the beam while keeping the spectrum unchanged to accommodate the detector dynamic range. The plates are positioned under a small angle to prevent a multiple reflection reaching the detector. The transmission of each plate has been experimentally determined by gradually inserting the plates to attenuate the THz signal strength. The THz beam is then spectrally characterized by a scanning Fourier transform spectrometer (Michelson interferometer), selected for its exceptionally wide supported bandwidth and absence of material limitations. The interferometer is designed to support a scan step as short as 300 nm, providing a reliable spectral resolution across the measured wavelengths $> 1.5$ μm (corresponding to $200$ THz). The generated THz beam diameter is in all our measurements $< 20$ mm and the whole system is designed to accommodate the diameter up to 25 mm, ensuring the collection of the whole generated THz beam and its delivery to the detector. The detector used in the measurements is a calibrated pyroelectric detector, specifically a WiredSense MPY-RS with a diamond window. The absolute calibration of the detector was performed by the manufacturer at 10.6 μm using a $CO_2$ nanosecond pulsed laser and at 1.55 μm using a pulsed laser diode, ensuring accurate absolute values within our spectral range of interest. Moreover, to cross-check the measured values, we performed additional measurements with a second, identical detector.

## Plasma defocusing effects

One of the most important factors reducing the intensity reached inside the generation medium is the defocusing caused by generated plasma, which increases the focal spot size compared to the vacuum measurement shown in Fig. 1b. Consequently, it is challenging to reach a very high intensity in high gas pressure due to the strong plasma defocusing, occurring at electron densities reaching $10^{20}$ cm$^{-3}$ [45].

The refraction of plasma formed by on-axis fast ionization becomes significant if the plasma length $L_{diff}$ required to form an on-axis phase shift of $\delta n k_0 L_{diff} = \pi/2$ is shorter than the Rayleigh range $z_R = \frac{\pi w_0^2}{\lambda}$ of the beam [46]. Considering the refractive index change in plasma given by $\delta n \approx \frac{n_e}{2n_c}$, where $n_e$ is the electron density and $n_c$ is the critical density, we obtain the condition

$$L_{diff} = \frac{\lambda}{2}\frac{n_c}{n_e} < \frac{4\lambda}{\pi} f_{\#}^2, \tag{1}$$

where $f_{\#}$ is the f-number of the driving beam. From the equation (1) we obtain the condition for the threshold electron density causing significant defocusing effects stated as

$$n_e > \frac{\pi}{8} n_c f_{\#}^{-2}. \tag{2}$$

Considering our experimental parameters, we obtain the limiting electron density

$$n_e \approx 2.5 \times 10^{19}\ \mathrm{cm}^{-3}. \tag{3}$$

By keeping the electron density low enough during the measurements (considering the pressure and the ionization degree) and not substantially exceeding this limit, we can ensure that the defocusing effects caused by the plasma do not alter the measured focal spot size (Fig. 1b) and, therefore, the calibrated laser intensity. We performed our measurements using sharp focusing ($f_{\#} = 5$) and with gas pressure limited to 200 mbar, maintaining the maximum theoretically reachable electron density for krypton ionized to $Kr^{10+}$ at $5 \times 10^{19}$ cm$^{-3}$.

# Ionization by a laser field.

For intensities of electric field defined by a Keldysh parameter $\gamma \ll 1$, tunneling ionization (TI) is the most suitable approximation. It can be described by the ionization rate $w_{ADK}$ introduced by Ammosov, Delone and Krainov (ADK) [47]

$$w_{ADK} = |C_{n^*l^*}|^2 G_{lm} I_p \left(\frac{2F_0}{F}\right)^{2n^*-|m|-1} e^{-\frac{2F_0}{3F}}, \tag{4}$$

where $n^* = \frac{Z_c}{\sqrt{2I_p}}$ and $l^* = n^* - 1$ are effective principal and orbital quantum numbers, $m$ is the magnetic quantum number, $|C_{n^*l^*}|^2 = \frac{2^{2n^*}}{n^*\Gamma(n^*+l^*+1)\Gamma(n^*-l^*)}$, $G_{lm} = \frac{(2l+1)(l+|m|)!}{2^{|m|}|m|!(l-|m|)!}$, $I_p$ is the ionization potential, $F_0 = (2I_p)^{3/2}$ is a Coulomb electric field, $F$ is the intensity of applied electric field and $Z_c$ is the charge of the ion. The ionization rate formula was later corrected for higher intensities compared to a barrier-suppression strength $F_b$ by Tong and Lin [48] or even higher intensities by Zhang, Lan and Lu [49]. For very high intensities in the far over-barrier ionization process [50], the ionization rates start to depend quadratically on intensity [51] followed by a linear region [52]. However, all correction parameters are known only for the neutrals or for the first ionized state of most usual atoms.

Therefore, to calculate ionization rates for higher ionization stages, we use the original ADK model. Even though the required peak intensities are well above the $F_b$ and thus at first sight seem to be outside of the formal validity range, we can still apply this simple model due to the long pulses used (40 fs). As the field slowly increases in intensity, the probability of ionization increases towards 1 and once the field reaches the intensity above the validity condition, the gas has already been fully ionized in a given ionization state. Therefore, further calculation for higher intensity is not necessary [53]. To calculate the ionization up to the desired intensities of $5 \times 10^{17}\ \mathrm{W/cm^2}$ we go up to $Kr^{10+}$, $Ne^{8+}$ and $He^{2+}$. Table 1 shows the electronic structure used for the calculation. During fast sequential ionization, in contrast to steady-state ionization, the quantum numbers of electrons in each orbital of subshells remain unchanged and the electron orbitals with $m = 0$ are ionized first followed by $m = \pm 1$ orbitals [53].

| Ionization degree | 0 | 1+ | 2+ | 3+ | 4+ | 5+ | 6+ | 7+ | 8+ | 9+ |
|---|---|---|---|---|---|---|---|---|---|---|
| Kr ([Ar] $3d^{10}4s^24p^6$) – *l* | 1 | 1 | 1 | 1 | 1 | 1 | 0 | 0 | 2 | 2 |
| Kr ([Ar] $3d^{10}4s^24p^6$) – *m* | 0 | 0 | -1 | -1 | 1 | 1 | 0 | 0 | 0 | 0 |
| Ne ([He] $2s^22p^6$) – *l* | 1 | 1 | 1 | 1 | 1 | 1 | 0 | 0 | | |
| Ne ([He] $2s^22p^6$) – *m* | 0 | 0 | -1 | -1 | 1 | 1 | 0 | 0 | | |
| He ($1s^2$) – *l* | 0 | 0 | | | | | | | | |
| He ($1s^2$) – *m* | 0 | 0 | | | | | | | | |

**Table 1. Electronic configurations.** Values of orbital quantum number *l* and magnetic quantum number *m* for krypton, neon and helium for used ionization degrees in the calculations of ionization.

# Plasma current model of THz generation

Once we obtained ionization rates for different ions and for different laser intensities, we calculate the temporal evolution of the electron density during the laser pulse. To get better insight into the whole process we treat the electron populations from different ionization stages separately considering sequential ionization $Kr \rightarrow Kr^{1+} \rightarrow Kr^{2+}$ ... . Furthermore, because of the particularly high intensities it is also necessary to consider the ground state depletion during ionization. The rate equations for electron density populations from different ionization states $N_e^k$ and densities of ions $N_g^k$ are defined as

$$\frac{dN_e^k(t)}{dt} = N_g^k(t)w^k(t) \tag{5}$$

$$N_g^k(t) = \begin{cases} N_e^{k-1}(t) - N_e^k(t), & \mathrm{k} > 1 \\ N_g^1(0) - N_e^{\mathrm{k}}(t), & \mathrm{k} = 1, \end{cases} \tag{6}$$

where $k$ states for the *k-th* ionization degree and $N_g^1$ is the density of neutrals. Not considering electron collisions, the radiated field contribution of electrons from the *k-th* ionization can be written from the corresponding current density $J^k(t)$ as [54]

$$E^k(t) \approx \frac{dJ^k(t)}{dt} = -\frac{e}{m_e} N_e^k(t)E(t). \tag{7}$$

The partial current density contribution at time $t$ from the electrons ionized at time $t_i$ can be therefore obtained from

$$dJ^k(t_i, t) = e\frac{dN_e^k(t_i)}{dt_i} v_e(t_i, t)dt_i = eN_g^k(t_i)w^k(t_i)v_e(t_i, t)dt_i, \tag{8}$$

where $e$ is electron charge with mass $m_e$ and $v_e(t_i, t)$ is the velocity of electrons at time $t$ that were ionized at time $t_i$ defined as

$$v_e(t_i, t) = -\frac{e}{m_e}\int_{t_i}^{\mathrm{t}} E(t')dt' \,. \tag{9}$$

In our case, we calculate the $E_{THz}(t)$ for a bi-chromatic driving laser field defined as $E(t) = E_{01}(t)\cos(\omega_1 t) + E_{02}(t)\cos(\omega_2 t + \varphi)$, where we consider $\omega_2 = 2\omega_1$, pulse duration $FWHM = 40$ fs , $\lambda_1 = 800$ nm , 5 % of second harmonic energy (corresponding to $E_{02} \approx 0.23E_{01}$) and phase delay $\varphi = {}^{\pi}/_{2}$.

For a more valuable insight into the temporal dependence of the contributions from individual ions, we removed the high-frequency oscillation part at the main frequency of 375 THz (800 nm) and two sub-harmonics of 187 THz (1600 nm) and 125 THz (2400 nm). By doing so, we can isolate the low-frequency, time-dependent current density contributions as the source term corresponding to the emitted THz field.

The THz spectra are subsequently obtained by the Fourier transform $S'_{THz}(\omega) = \mathcal{F}\big(E_{THz}(t)\big)$.

So far, the propagation of the generated THz field in the plasma medium has not been considered. To address the plasma frequency propagation cut-off of the resulting spectra of $E_{THz}(t)$, we employ the spectrally dependent absorption factor $a(\omega) = e^{-k_i(\omega)z}$, where $z$ is the propagation distance in the order of few µm and $k_i$ is the imaginary component of k-vector

$$k(\omega) = \frac{\omega}{c_0}\sqrt{1 - \frac{\omega_p^2}{\omega^2}}. \tag{10}$$

Here, $\omega_p$ is the plasma frequency dependent on the total electron density $N_e = \sum_k N_e^k$

$$\omega_p = \sqrt{\frac{N_e e^2}{\varepsilon_0 m_e}}. \tag{11}$$

The final spectra are therefore obtained applying the frequency dependent absorption induced by plasma

$$S_{THz}(\omega) = a(\omega)S'_{THz}(\omega). \tag{12}$$

# RESULTS

## Dependence of THz signal strength on laser intensity

We observed a high conversion efficiency of THz generation with the significantly increased driving laser intensity (Fig. 2). We link this to the high ionization degree of plasma. To demonstrate the influence of multiple ionization on the efficiency of THz generation, we show the example of krypton gas that offers the possibility of reaching wealth of higher ionization states (up to $Kr^{10+}$) using achievable intensity ($4 \times 10^{17}$ W/cm$^2$). In Fig. 2, we present the experimental results of the influence of the krypton states between $Kr^{5+}$ and $Kr^{10+}$ on the generation at high driving laser intensity highlighting the major contribution of higher ionization states on the resulting THz energy.

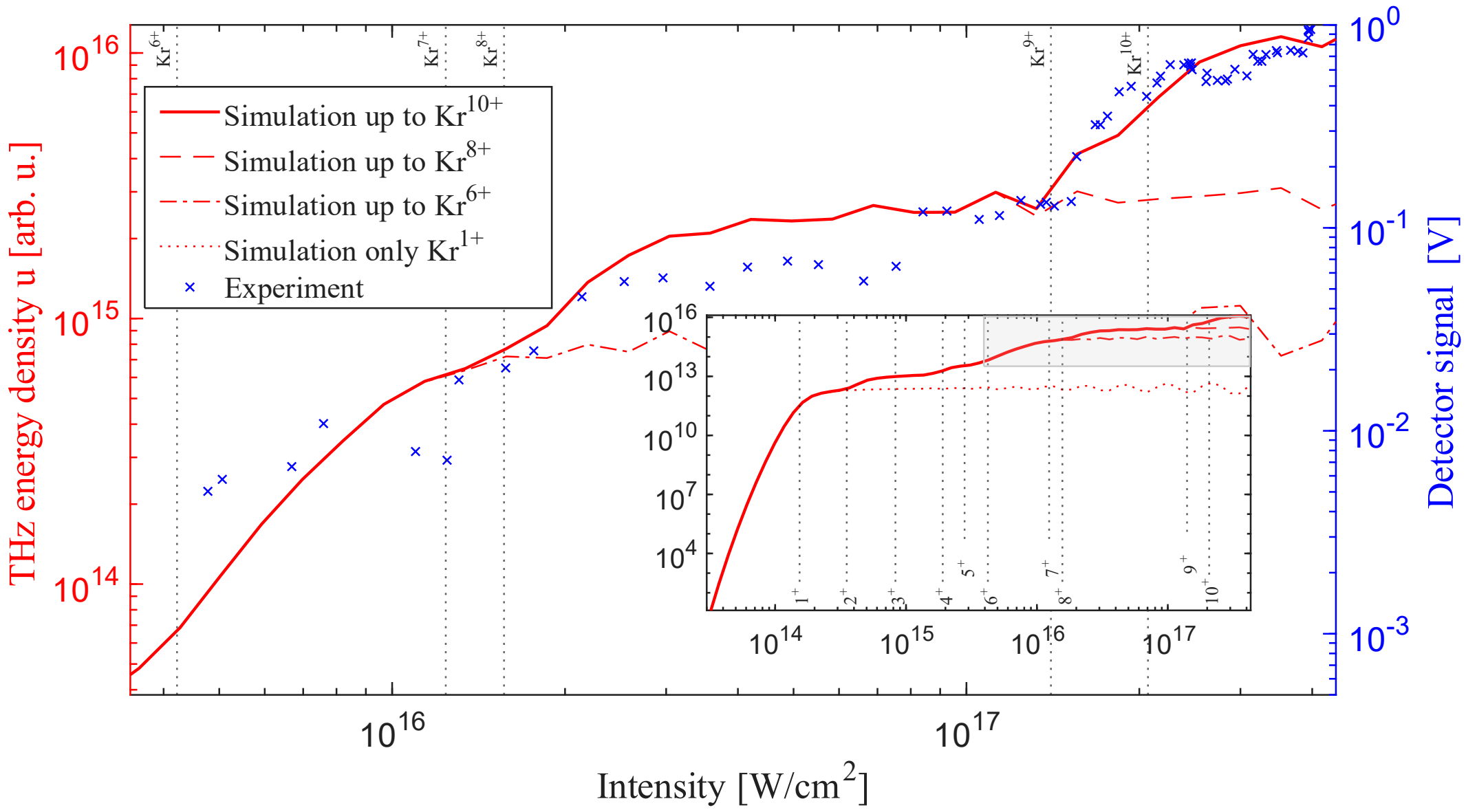

**Fig. 2 THz generation dependence on driving laser intensity employing various ionization states of the gas.** Measurement of the total energy of the THz radiation generated in krypton at 200 mbar for laser intensities between $5 \times 10^{15}$ and $5 \times 10^{17}$ W/cm$^2$. Measured data are shown in blue crosses and simulation results are shown as red lines. The solid red line shows the simulation with all ionization states, and the red dashed lines show simulations with ionizations up to $Kr^{1+}$, $Kr^{6+}$, and $Kr^{8+}$, respectively. An inset shows an extended range of intensities for simulated results between $3 \times 10^{13}$ and $5 \times 10^{17}$ W/cm$^2$. A gray shaded area in the inset shows the location of the main plot with experimental data. Dashed vertical lines represent the critical threshold intensity corresponding to the barrier-suppression-ionization field $F_{BSI} = I_p^2/4Z_c$.

In Fig. 2, a “staircase-like” structure of the THz pulse energy is visible in both experimental and calculated results. Each stair is caused by a higher ionization state contributing to the generation process. In regions with larger intensity separation between consecutive ionization thresholds, the THz energy rises exponentially once ionization occurs and quickly saturates when the ions reach the next ionization state. This behavior averages out to a linear rise in regions where consecutive ionizations occur at similar intensities, such as between $Kr^{4+}$ and $Kr^{8+}$.

The substantial agreement between experimental results and theory regarding this “staircase-like” structure demonstrates that the effect is driven by the gradual involvement of individual ionization states, rather than volumetric effects or plasma channel formation. To further confirm that this structure is solely due to consecutive ionizations, we performed a simulation considering only $Kr^{1+}$, as well as simulations including only ionization states up to $Kr^{6+}$ and $Kr^{8+}$, respectively (see red dashed lines in Fig. 2). One can see that when higher ionization states are omitted, the THz signal quickly saturates, such that increasing driver intensity yields no further signal enhancement.

To simulate the THz generation, we employed a photocurrent model based on the multiple ionization states of the gas (see Methods). The qualitative comparison of the measured results together with simulations can be seen in Fig. 2. First, the increase of the THz pulse energy scales linearly with the laser intensity. To show the importance of multiple ionization we note that the obtained THz energy is higher by 4 orders of magnitude compared to a single ionized state, while the laser intensity needed to reach this regime is 3 orders of magnitude higher, as it is visible in the inset of Fig. 2. Although the qualitative agreement is particularly good, one can note a small quantitative disagreement comparing experimental and simulated data, which might be due to propagation and volumetric effects not included in our simulations.

## Impact of high laser intensity on THz spectrum

Besides the overall THz pulse energy, the higher ionization degrees also strongly impact the THz spectrum. One of the general characteristics of the THz spectrum at laser intensities $\approx 10^{14}\ \mathrm{W/cm^2}$ is broadening towards high THz frequencies with increasing laser intensity [21, 22]. However, in rare THz studies reaching even higher intensities [22, 37, 41], the observed phenomena is not only the broadening of the spectrum but also a shift to higher frequencies (few tens of THz) with suppressed signal in low frequencies that is in contradiction with broadening derived from the state-of-the-art (single ionization) photocurrent model. This phenomenon is usually not discussed in detail or attributed to the spectral broadening seen in simulations for intensities $\approx 10^{14}\ \mathrm{W/cm^2}$.

The spectral shift towards higher frequencies can be seen from our experimental observations for krypton in Fig. 3a and for neon in Fig. 3b. Note a clear displacement of the main spectral peak to frequencies around 60-70 THz with higher driving intensities. The low frequency part is clearly suppressed, which is attributed to the plasma reabsorption arising from the increased electron density. From the established photocurrent model without plasma reabsorption, the reduced conversion efficiency of this part of the spectrum cannot be otherwise explained. The suppression of the low frequency part serves therefore as an in-situ measurement of the plasma density in the order of $\approx 10^{19}\ \mathrm{cm^{-3}}$ ($\omega_p \approx$ a few tens of THz), confirming the laser intensity reached inside the medium.

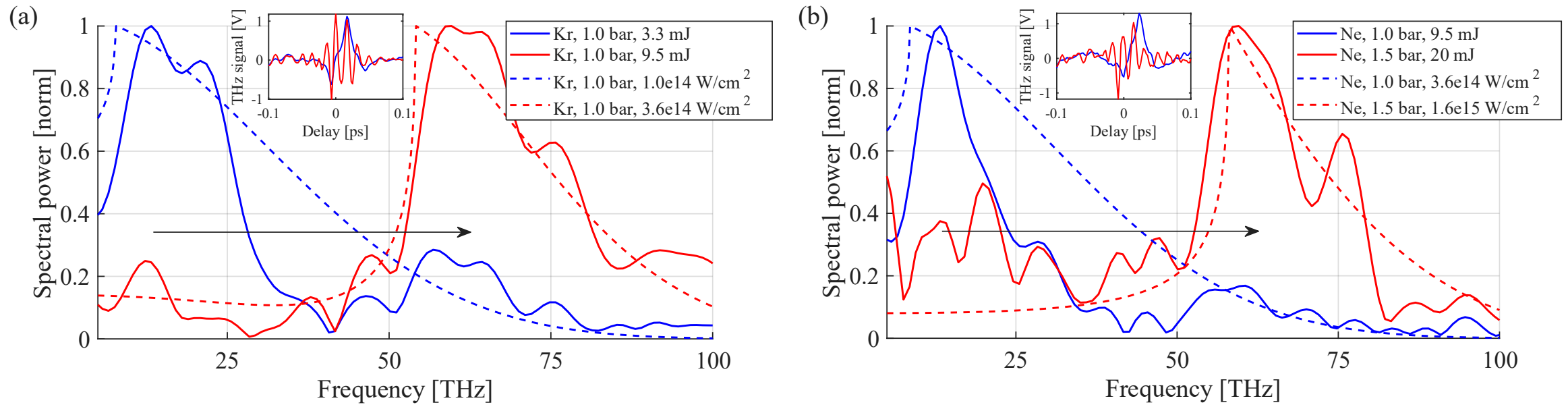

**Fig. 3 THz spectral dependence on driving laser intensity. (a)** Spectral measurement for two laser energies, 3.3 mJ and 9.5 mJ, for krypton at 1 bar pressure. **(b)** Spectral measurement for 9.5 mJ energy in 1 bar of neon and comparison with increased energy to 20 mJ and pressure to 1.5 bar pressure. In both cases, the simulation data are shown with dashed lines and insets show measured corresponding autocorrelation traces. The black arrows highlight the spectral shift to higher frequences due to increased electron density.

Even though the simulation results show excellent agreement with the measured data, the maximum intensity cannot be directly linked to the laser pulse energy because of defocusing at higher gas pressures ($> 200$ mbar), discussed in the section Plasma defocusing effects. Additionally, in our experimental setup, the THz diagnostic is performed in ambient air, which necessitates accounting for water absorption, occurring mostly between 45-50 THz [55].

# Influence of gas ionization potential on generated THz fields

It is important to note, however, that the spectrum is not determined solely by the driving field intensity. The dependence on gas ionization potential is shown by comparing Fig. 3a (red line) for krypton and Fig. 3b (blue line) for neon. Despite being measured under the same conditions (1 bar, 9.5 mJ), the two gases produced significantly different spectrum. As neon has higher ionization potential, less electron density is created than in the krypton case. Furthermore, a comparable electron density leading to a similar THz spectrum of neon (Fig. 3b, blue line) can be achieved by reducing the intensity in the krypton case (Fig. 3a, blue line). This is also confirmed by the simulations (dashed lines) done for corresponding pressure, target gas, and the driving laser intensity. These results indicate that the generated THz spectrum depends not exclusively on the intensity, but also on the ionization dynamics and resulting electron density. Thus, it is connected to all three parameters: laser intensity, ionization potential of the gas and atomic density. In conclusion, both experimental and simulation results suggest that the THz generation is driven by time-dependent electron density and driving electric field of the laser (equation (7)).

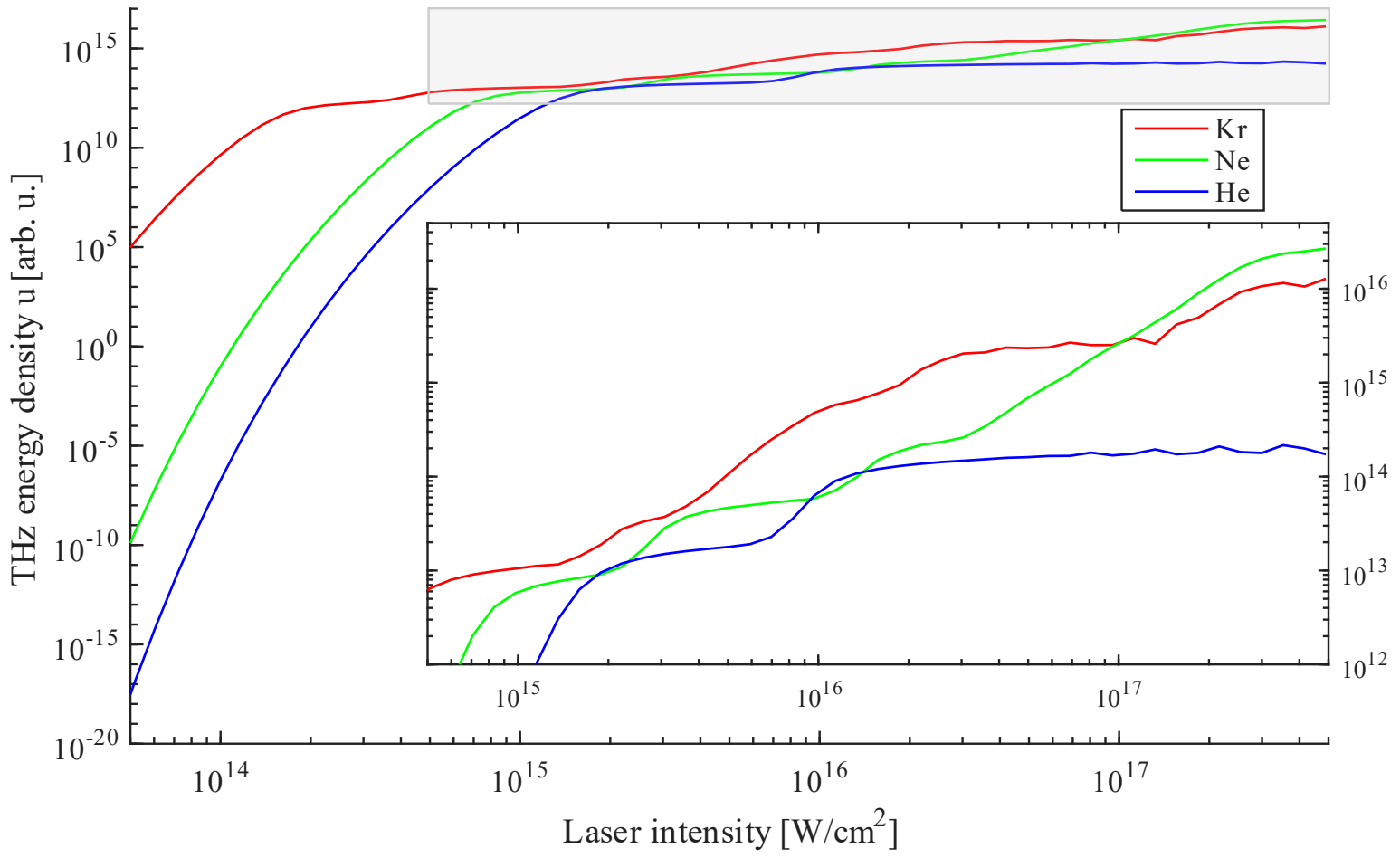

**Fig. 4 Dependence of the THz generation efficiency on laser driving intensity.** Simulation results on the dependence of the generated THz energy on the laser intensity between $3 \times 10^{13}$ and $5 \times 10^{17}\ \mathrm{W/cm^2}$ for different gases at 200 mbar. An inset shows a zoom of the main graph marked by a gray shaded area.

To give additional insight into the dependence of generated THz energy on the ionization potential of the gas species (through the time-dependent electron density), we show the simulated results of the THz pulse energy for three gases at the same gas pressure (Fig. 4). If we look at helium that has two electrons (blue line in Fig. 4), we can see the stair-like pattern composed of two steps only, each corresponding to one ionization degree. In the intensity region above $2 \times 10^{16}\ \mathrm{W/cm^2}$, the helium is fully ionized and thus the generated THz energy stops increasing.

A similar behavior can be observed when comparing the intensity-dependent spectra for krypton and helium in Fig. 5a and Fig. 5b (note the same range of both axes and color scale). For krypton, not only is the conversion efficiency higher, but the spectral peak also shifts towards higher frequencies while lower frequencies are attenuated due to plasma reabsorption. For helium (Fig. 5b), the low frequency limit is unaffected by laser intensities above $2 \times 10^{16}\ \mathrm{W/cm^2}$. As shown in Fig. 4, this corresponds to a plasma absorption dependent on the electron density limited by a full ionization of helium (2 electrons) and a change in pressure of the gas. In Fig. 6a we can see the comparison of the experimentally measured spectra for three gases at the same conditions. The spectral shift to higher frequencies is clearly caused by the rising electron density due to the lower ionization potential mostly visible for krypton.

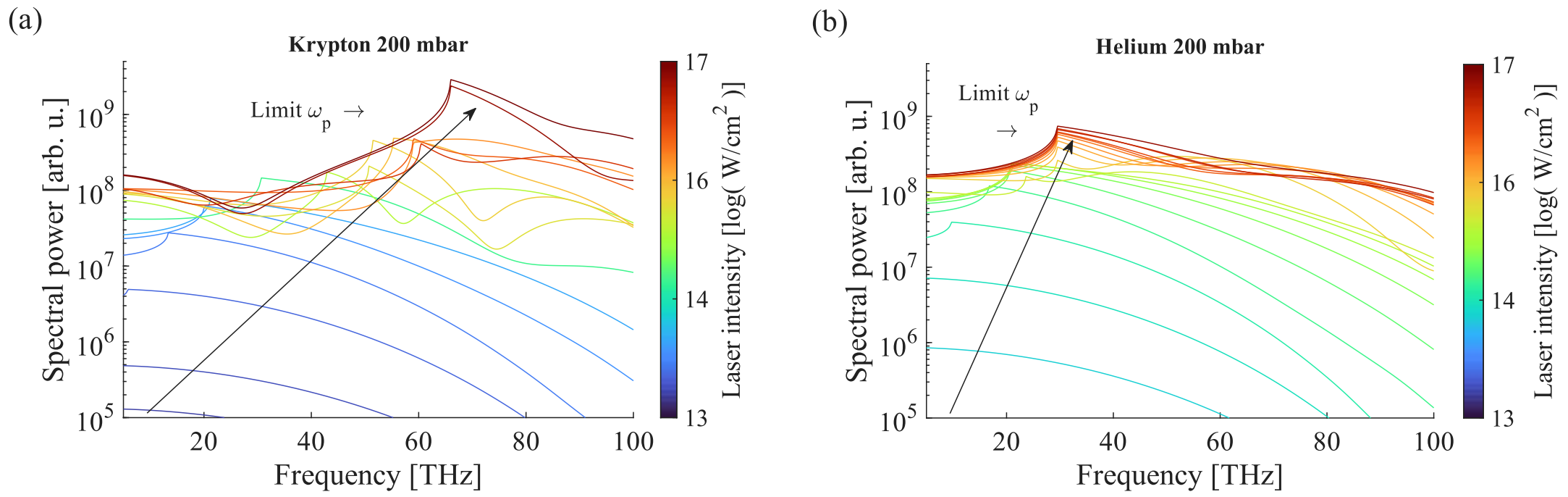

**Fig. 5 Dependence of the THz generated spectrum on laser driving intensity.** Simulation results on the evolution of the generated THz spectral power on the laser intensity between $3 \times 10^{13}$ and $5 \times 10^{17}$ W/cm$^2$ for **(a)** krypton and **(b)** helium, both at 200 mbar. The black arrows highlight the spectral shift to higher frequences due to increased electron density.

# Influence of ionization rate on generated THz fields

Furthermore, in Fig. 4 we can note that for certain driving intensities, the yield from gas with lower number of electrons can counterintuitively exceed the one from gas with a higher number of electrons. Helium overtakes neon at intensity of $\approx 1 \times 10^{16}\ \mathrm{W/cm^2}$ and neon yields more than krypton for intensities $> 1 \times 10^{17}\ \mathrm{W/cm^2}$. This demonstrates that besides the total electron density, the ionization rate is a crucial factor. In our case, for intensities $> 1 \times 10^{17}\ \mathrm{W/cm^2}$, $Kr^{9+}$ and $Kr^{10+}$ have a different electron shell involved as compared to $Ne^{7+}$ and $Ne^{8+}$ (3d vs. 2s, see Table 1) with different ionization rates.

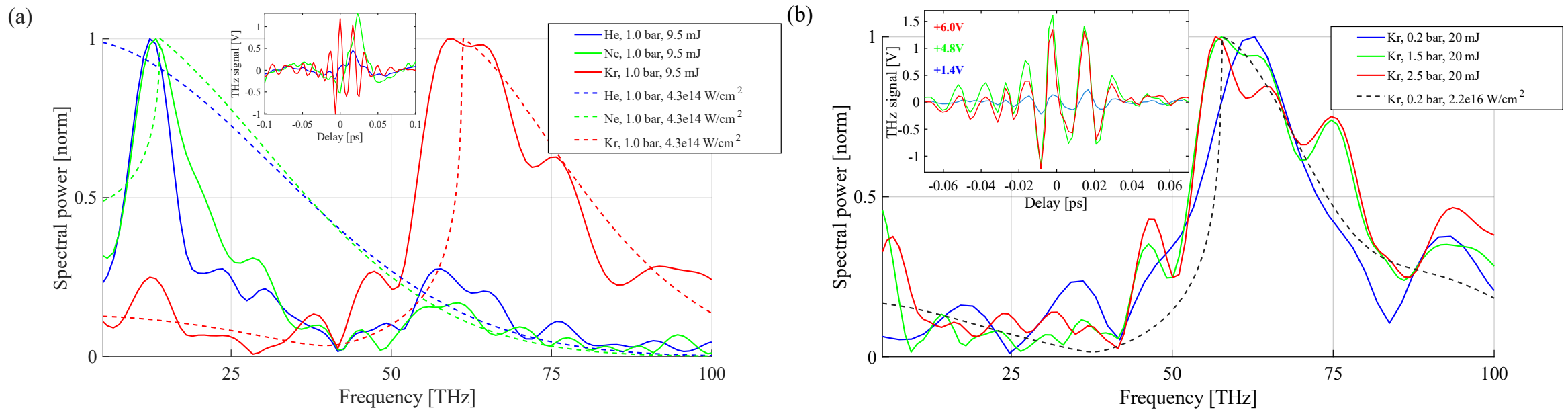

**Fig. 6 Dependence of the THz spectrum on plasma electron density. (a)** Experimental and simulation results on THz generated spectra dependent on the ionization potential of the three gases (He, Ne, Kr) in the same conditions of 1 bar and 9.5 mJ driving pulse energy. **(b)** Experimental results of the THz spectrum dependence on the krypton gas pressure in the overdriven regime with maximal plasma density reached. In the inset, the absolute THz signal values (subtracted baselines) are noted in corresponding colors. In both, the simulation results are shown by dashed lines in corresponding colors and insets are showing corresponding experimentally measured autocorrelation traces.

# In situ probing of plasma defocusing

To prove that the low THz frequency limit given by a plasma frequency is dependent only on the electron density, we measured the THz spectra for krypton (a gas with low ionization potential and many reachable degrees of ionization) at a very high laser energy (20 mJ). To demonstrate the electron density clamping around $5 \times 10^{19}\ \mathrm{cm^{-3}}$ $\left(\omega_p/2\pi \approx 64\ \mathrm{THz}\right)$ due to plasma defocusing, we show that above certain driving laser intensity and gas pressure, the THz spectrum does not change with further pressure increase, once this limiting electron density has been reached. Fig. 6b shows the identical spectral shape for gas pressures of 0.2, 1.5, and 2.5 bars, driven by the same laser energy of 20 mJ. From the autocorrelation traces in the inset, we can see a significantly higher THz field strength for higher pressures. Note that the traces are plotted with subtracted baselines (absolute THz signal strength), whose values are stated in the inset in corresponding colors. While the THz spectra remain unchanged due to plasma defocusing-induced intensity clipping, the increase in the overall efficiency is attributed to volumetric effects.

Moreover, for krypton at 2.5 bars and 20 mJ driving laser energy, we measured the generated THz energy to be 220 µJ (average of 300 consecutive shots measured after 6 thin HRFZ-Si attenuating plates and 1 HRFZ-Si window – see Fig. 1a) giving a conversion efficiency of 1.1 %, one of the highest reached with an 800 nm driving laser to our knowledge [25, 28, 41].

# Temporal profile of a THz pulse

One of the important aspects of THz generation in the high-intensity regime, usually omitted in literature, is the temporal shape of the THz pulses. As we can see in the insets of experimentally measured spectra (e.g. Fig. 6b), the autocorrelation traces are becoming a multicycle for high intensities in contrast to single-cycle THz wave generated by state-of-the-art low intensity THz sources. The same effect translated into the spectral shape can be visible from the simulations presented in Fig. 5a for krypton, where for high driving intensities the generated spectrum has a complex structure compared to helium in Fig. 5b.

The simulated temporal profile of the THz pulse generated by a high-intensity driving laser (Fig. 7a) shows that the pulse shape can have a nontrivial envelope. Note that, the $E_{THz}(t)$ is only the source part of the emitted THz pulse by the current density $J(t)$ (shown in Fig. 7c), as we are not solving the recombination part of the electronic current after the driving laser pulse ends (the current density in Fig. 7c and Fig. 7d after the pulse stays in stable DC value). The reason is, that the recombination is significantly slower than the ionization and thus does not contribute to the THz generation.

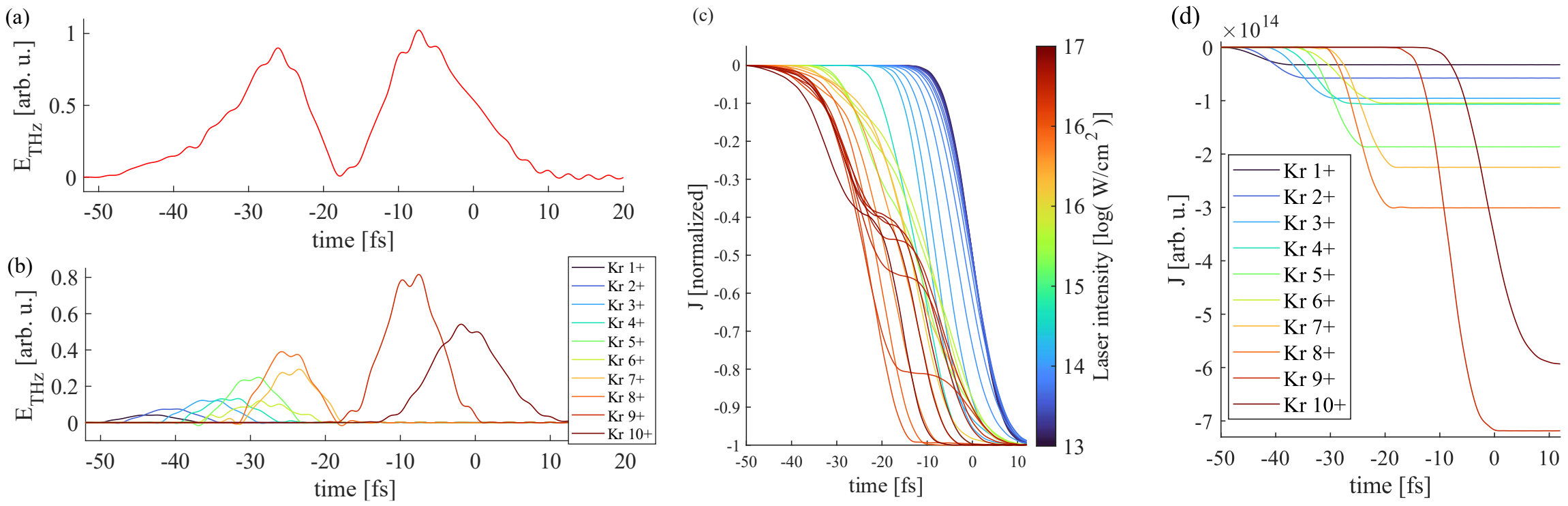

**Fig. 7 Temporal characteristics of THz emission from multiply ionized plasma. (a)** Emitted THz pulse for a laser intensity of $2.2 \times 10^{17}\ \mathrm{W/cm^2}$ from krypton at 200 mbar. **(b)** Corresponding temporal THz emission contributions from different ionization states (their sum is shown in **(a)**). **(c)** Normalized temporal dependence of the total current density $J$ through the driving laser pulse for different laser intensities between $3 \times 10^{13}$ and $5 \times 10^{17}\ \mathrm{W/cm^2}$ shown in color code. $J$ is shown without the high frequency oscillating component. **(d)** Temporal dependence of the current density contributions from different ionization states for laser intensity of $2.2 \times 10^{17}\ \mathrm{W/cm^2}$ in krypton at 200 mbar (time derivate of each contribution corresponds to **(b)**).

We attribute this complex temporal THz pulse profile to the temporal evolution of different electron populations (from different ionization states) within the pulse duration of the driving laser field. As the electron density is time dependent, the electric current $J$ becomes time dependent as well, as shown in Fig. 7c. First, if we look at the current density evolution within the laser pulse duration, dependent on the driving laser intensity shown in Fig. 7c, we can note that it forms a double-fold structure for intensities above $\approx 1 \times 10^{16}\ \mathrm{W/cm^2}$. The effect happens for each driving field intensity, where the successive ionization occurs (see the stair-like structure in Fig. 2 or Fig. 4). Therefore, this double-folded structure can be understood by decomposing it into the contributions from individual ionization degrees. Fig. 7d shows the evolution of the current densities for different charge states of 200 mbar of krypton at the laser intensity of $2.2 \times 10^{17}\ \mathrm{W/cm^2}$. The sum of these contributions corresponds to the dark red curve in Fig. 7c.

The contributions to the emitted THz field from all involved ions are resolved in time in Fig. 7b. The figure shows that the higher ionization states have a much larger impact on the amplitude of the emitted field and therefore significantly increase the total conversion efficiency. This is because the ionization of the higher orders occurs around the driving pulse peak, where the initial asymmetric push on the electron at the time of ionization maximizes.

The variation in the current density throughout the entire driving pulse duration is linked to the pulse duration of the emitted THz field. The pulse envelope in Fig. 7a spans $\approx 40$ fs, being the sum of the contributions from Fig. 7b, which are derived from currents shown in Fig. 7d. If we consider its spectrum centered at high frequencies (Fig. 6b), an autocorrelation of such a temporal structure clearly produces a multi-cycle pulse, in full agreement with experimental data shown in the insets (e.g., Fig. 6b).

# DISCUSSION

We have developed and validated a photocurrent model employing high ionization states of noble gases. We have shown that our model fills the gap of understanding the THz generation at high laser intensities spanning up to $5 \times 10^{17}\ \mathrm{W/cm^2}$ where high ionization states such as $Kr^{10+}$ and $Ne^{8+}$ occur. We have provided the comparison with experimental measurements showing excellent agreement in both spectral and conversion efficiency behaviors. Particularly, we addressed and provided a full description of the spectral shift – the feature that was already reported but never properly understood. Here, in agreement with the experimental results, we provide the explanation for the spectral shift and the attenuation of low frequency component of the generated THz radiation. We attribute these effects to the plasma frequency determined by the electron density.

In our experiments, we achieved conversion efficiency exceeding 1 % with 800 nm driving laser wavelength. This high efficiency of the THz generation results from maximizing the gain in the electron density via high intensity of the 800 nm pulse. The achieved efficiency is comparable to the yields reported for mid-IR lasers [25] and significantly exceeds the ones reported for Ti:sapphire laser systems [27, 29, 37].

Moreover, our plasma current model, including all ionization states of the gas, reveals the temporal aspects of THz emission from the understanding of ionization evolution throughout the whole envelope of the driving laser pulse. This model explains experimentally observed phenomena of multicycle THz field emission from laser plasma in the high-intensity regime that has remained unexplored until now.

We believe our work shows a direction for further optimization and upscaling of the THz generation in two-color laser plasma, with a promising outlook of multi-percent conversion efficiency with NIR laser systems and potentially even higher, if applied on mid-IR laser sources opening the prospects of new class of nonlinear THz applications.

**Acknowledgments:**
This work was carried out at the ELI Beamlines Facility, a European user facility operated by the Extreme Light Infrastructure ERIC. We thank the facility staff for their assistance and their support.

**Author contributions:**
L.J. and O.H. conceived the project and designed and built the experimental setup; L.J. developed the theory and wrote the manuscript; D.S., C.J. and M.S. performed the experiment under the supervision of L.J.; D.S., C.J. and L.J. analyzed the data; D.S. and J.N. contributed to the development of the theory. All authors reviewed the manuscript.

**Funding:** L.J. and J.N. acknowledge the support from the Czech Science Foundation Grant No. 25-17853S.

**Competing interests:** The authors declare that there is no conflict of interest regarding the publication of this article.

**DATA AVAILABILITY**
All relevant data is available on request from the corresponding author.

## References

[1] Tonouchi M., "Cutting-edge terahertz technology," *Nature Photon 1(2),* p. 97–105, 2007.

[2] Zhang X. C., Shkurinov A., Zhang Y., "Extreme terahertz science," *Nature Photon 11(1),* p. 16–18, 2017.

[3] Salén P., Basini M., Bonetti S., Hebling J., Krasilnikov M., Nikitin A. Y., Shamuilov G., Tibai Z., Zhaunerchyk V., Goryashko V., "Matter manipulation with extreme terahertz light: Progress in the enabling THz technology," *Physics Reports, 836,* pp. 1-74, 2019.

[4] Vicario C., Shalaby M., Hauri C. P., "Subcycle Extreme Nonlinearities in GaP Induced by an Ultrastrong Terahertz Field," *Phys. Rev. Lett., 118(8), 083901,* 2017.

[5] Shen Y., Watanabe T., Arena D. A., Kao C. C., Murphy J. B., Tsang T. Y., Wang X. J., Carr G. L., "Nonlinear Cross-Phase Modulation with Intense Single-Cycle Terahertz Pulses," *Phys. Rev. Lett., 99(4), 043901,* 2007.

[6] Shen Y., Carr G. L.,Murphy J. B., Tsang T. Y., Wang X., Yang X., "Electro-optic time lensing with an intense single-cycle terahertz pulse," *Phys. Rev. A 81, 053835,* 2010.

[7] Matsunaga R., Hamada Y. I., Makise K., Uzawa Y., Terai H., Wang Z., Shimano R., "Higgs Amplitude Mode in the BCS Superconductors Nb1-x TixN Induced by Terahertz Pulse Excitation," *Phys. Rev. Lett. 111, 057002,* 2013.

[8] Mankowsky R., Subedi A., Först M., Mariager S. O., Chollet M., Lemke H. T., Robinson J. S., Glownie J. M., Minitti M. P., Frano A., et al., "Nonlinear lattice dynamics as a basis for enhanced superconductivity in YBa2Cu3O6.5," *Nature, 516(7529),* p. 71–73, 2014.

[9] Hort O., Dubrouil A., Khokhlova M. A., Descamps D., Petit S., Burgy F., Mével E., Constant E., Strelkov V. V., "High-order parametric generation of coherent XUV radiation," *Opt. Express, 29(4),* pp. 5982-5992, 2021.

[10] Birulia V. A., Khokhlova M. A., Strelkov V. V., "Generation of attosecond pulses with a controllable carrier-envelope phase via high-order frequency mixing," *Phys. Rev. A, 106(2), 023514,* 2022.

[11] Birulia V. A., Khokhlova M. A., Strelkov V. V., "Macroscopic effects in generation of attosecond XUV pulses via high-order frequency mixing in gases and plasma," *New J. Phys., 26(2), 023005,* 2024.

[12] Kovács K., Balogh E., Hebling J., Toşa V., Varjú K., "Quasi-Phase-Matching High-Harmonic Radiation Using Chirped THz Pulses," *Phys. Rev. Lett. 108(19), 193903,* 2012.

[13] Tóth G., Tibai Z., Sharma A., Fülöp J. A., Hebling J., "Single-cycle attosecond pulses by Thomson backscattering of terahertz pulses," *J. Opt. Soc. Am. B, 35(5),* pp. A103-A109, 2018.

[14] Balogh E., Kovacs K., Dombi P., Fulop J. A., Farkas G., Hebling J., Tosa V., Varju K., "Single attosecond pulse from terahertz-assisted high-order harmonic generation," *Phys. Rev. A, 84(2), 023806,* 2011.

[15] Zhang D., Fallahi A., Hemmer M., Wu X., Fakhari M., Hua Y., Cankaya H., Calendron A.-L., Zapata L. E., Matlis N. H., et al., "Segmented terahertz electron accelerator and manipulator (STEAM)," *Nature Photon 12(6),* p. 336–342, 2018.

[16] Huang R. W., Fallahi A., Wu X., Cankaya H., Calendron A. L., Ravi K., Zhang D., Nanni E. A., Hong K.-H., Kärtner F. X., "Terahertz-driven, all-optical electron gun," *Optica 3(11),* pp. 1209-1212, 2016.

[17] Nanni E. A., Huang W. R., Hong K. H., Ravi K., Fallahi A., Moriena G., Dwayne M. RJ., Kärtner F. X., "Terahertz-driven linear electron acceleration," *Nat Commun, 6(1), 8486,* 2015.

[18] Tóth G., Polónyi G., Hebling J., "Tilted pulse front pumping techniques for efficient terahertz pulse generation," *Light Sci Appl, 12(1), 256,* 2023.

[19] Fülöp J. A., Ollmann Z., Lombosi Cs S., Skrobol C., Klingebiel S., Pálfalvi L., Krausz F., Karsch S., Hebling J., "Efficient generation of THz pulses with 0.4 mJ energy," *Opt. Express 22(17),* pp. 20155-20163, 2014.

[20] Vicario C., Ovchinnikov A. V., Ashitkov S. I., Agranat M. B., Fortov V. E., Hauri C. P., "Generation of 0.9-mJ THz pulses in DSTMS pumped by a Cr:Mg2SiO4 laser," *Opt. Lett., 39(23),* pp. 6632-6635, 2014.

[21] Xie X., Dai J., Zhang X. C., "Coherent Control of THz Wave Generation in Ambient Air," *Phys. Rev. Lett., 96(7), 075005,* 2006.

[22] Kim K. Y., Taylor A. J., Glownia J. H., Rodriguez G., "Coherent control of terahertz supercontinuum generation in ultrafast laser–gas interactions.," *Nature Photon 2(10),* p. 605–609, 2008.

[23] Mitrofanov A. V., Sidorov-Biryukov D. A., Nazarov M. M., Voronin A. A., Rozhko M. V., Shutov A. D., Ryabchuk S. V., Serebryannikov E. E., Fedotov A. B., Zheltikov A. M., "Ultraviolet-to-millimeter-band supercontinua driven by ultrashort mid-infrared laser pulses," *Optica 7(1),* pp. 15-19, 2020.

[24] Kuk D., Yoo Y. J., Rosenthal E. W., Jhajj N., Milchberg H. M., Kim K. Y., "Generation of scalable terahertz radiation from cylindrically focused two-color laser pulses in air," *Appl. Phys. Lett., 108(12), 121106,* 2016.

[25] Koulouklidis A. D., Gollner C., Shumakova V., Fedorov V. Y., Pugžlys A., Baltuška A., Tzortzakis S., "Observation of extremely efficient terahertz generation from mid-infrared two-color laser filaments," *Nat Commun., 11(1), 292,* 2020.

[26] Jang D., Schwartz R. M., Woodbury D., Griff-McMahon J., Younis A. H., Milchberg H. M., Kim K. Y., "Efficient terahertz and Brunel harmonic generation from air plasma via mid-infrared coherent control," *Optica 6(10),* pp. 1338-1341, 2019.

[27] Fedorov V. Y., Tzortzakis S., "Extreme THz fields from two-color filamentation of midinfrared laser pulses," *Phys. Rev. A, 97(6), 063842,* 2018.

[28] Clerici M., Peccianti M., Schmidt B. E., Caspani L., Shalaby M., Giguère M., Lotti A., Couairon A., Légaré F., Ozaki T., et al., "Wavelength Scaling of Terahertz Generation by Gas Ionization," *Phys. Rev. Lett. 110(25), 253901,* 2013.

[29] Fu B., Zhang Z, Liu Z., Wang K., Song L., Li W., Kostin V. A., Silaev A. A., Zhuang L.,Lu C., et al., "Efficient terahertz radiation generation from air plasma with mid- and near-infrared three-color femtosecond pulses," *Opt. Express 33(11),* pp. 22610-22631, 2025.

[30] Veisz L., Fischer P., Vardast S., Schnur F., Muschet A., De Andres A., Kaniyeri S., Li H., Salh R., Ferencz K., et al., "Waveform-controlled field synthesis of sub-two-cycle pulses at the 100 TW peak power level," *Nat. Photon., 19(9),* pp. 1013-1019, 2025.

[31] Viotti A. L., Seidel M., Escoto E., Rajhans S., Leemans W. P., Hartl I., Heyl C. M., "Multi-pass cells for post-compression of ultrashort laser pulses," *Optica 9(2),* pp. 197-216, 2022.

[32] Toth S., Stanislauskas T., Balciunas I., Budriunas R., Adamonis J., Danilevicius R., Viskontas K., Lengvinas D., Veitas G., Gadonas D., et al., "SYLOS lasers – the frontier of few-cycle, multi-TW, kHz lasers for ultrafast applications at extreme light infrastructure attosecond light pulse source," *J. Phys. Photonics, 2(4), 045003,* 2020.

[33] Antipenkov R., Novák J., Erdman E., Boge R., Szuba W., Tykalewicz B., Mazurek P., Lazzarini C., Grittani G., Bakule P., et al., "L1 Allegra laser at ELI Beamlines facility as a driver for electron acceleration at 1 kHz repetition rate," *Proc. SPIE PC12577,* 2023.

[34] Buldt J., Stark H., Müller M., Grebing C., Jauregui C., Limpert J., "Gas-plasma-based generation of broadband terahertz radiation with 640 mW average power," *Opt. Lett., 46(20),* pp. 5256-5259, 2021.

[35] Kim K. Y., Glownia J. H., Taylor A. J., Rodriguez G., "Terahertz emission from ultrafast ionizing air in symmetry-broken laser fields," *Opt. Express 15(8),* pp. 4577-4584, 2007.

[36] Zhang Z., Liu Z., Wang S., Lu C., Fan Z., Kostin V. A., Liu Y., "Spectral tailoring of the terahertz radiation from air plasma excited by two-color femtosecond pulses," *Appl. Phys. Lett. 123(3), 031108,* 2023.

[37] Zhao J., Zhang Y., Gao Y., Li M., Liu X., Liu W., Yan T.M., Jiang Y., "Temporal-spatial manipulation of bi-focal bi-chromatic fields for terahertz radiations," *Commun Phys, 7(1), 408,* 2024.

[38] Simpson T. T., Pigeon J. J., Ambat M. V., Miller K. G., Ramsey D., Weichman K., Froula D. H., Palastro J. P., "Spatiotemporal control of two-color terahertz generation," *Phys. Rev. Research. 6, 013041,* 2024.

[39] Liu K., Koulouklidis A. D., Papazoglou D. G., Tzortzakis S., Zhang, X. C., "Enhanced terahertz wave emission from air-plasma tailored by abruptly autofocusing laser beams," *Optica 3(6),* pp. 605-608, 2016.

[40] Zhang Z., Chen Y., Cui S., He F., Chen M., Zhang Z., Yu J., Chen L., Sheng Z., Zhang J., "Manipulation of polarizations for broadband terahertz waves emitted from laser plasma filaments," *Nature Photon., 12(9),* p. 554–559, 2018.

[41] Oh T. I., You Y. S., Jhajj N., Rosenthal E. W., Milchberg H. M., Kim K. Y., "Intense terahertz generation in two-color laser filamentation: energy scaling with terawatt laser systems," *New J. Phys., 15(7) 075002,* 2013.

[42] Fedorov Y. V., Koulouklidis A. D., Tzortzakis S., "THz generation by two-color femtosecond filaments with complex polarization states: four-wave mixing versus photocurrent contributions," *Plasma Phys. Control. Fusion, 59(1), 014025,* 2017.

[43] Koulouklidis A. D., Fedorov V. Y., Tzortzakis S., "Spectral bandwidth scaling laws and reconstruction of THz wave packets generated from two-color laser plasma filaments," *Phys. Rev. A, 93(3), 033844,* 2016.

[44] Kim, K. Y., "Generation of coherent terahertz radiation in ultrafast laser-gas interactions," *Phys. Plasmas, 16(5), 056706,* 2009.

[45] Monot P., Auguste T., Lompré L. A., Mainfray G., Manus C., "Focusing limits of a terawatt laser in an underdense plasma," *J. Opt. Soc. Am. B, 9(9),* pp. 1579-1584, 1992.

[46] Rankin R., Capjack C. E., Burnett N. H., Corkum P. B., "Refraction effects associated with multiphoton ionization and ultrashort-pulse laser propagation in plasma waveguides," *Opt. Lett. 16,* pp. 835-837, 1991.

[47] Ammosov M. V., Delone N. B., Krainov V.P., "Tunnelling ionization of complex atoms and of atomic ions in an alternating electromagnetic field.," *Sov. Phys. JETP 64(6), 1191,* 1986.

[48] Tong X. M., Lin C. D., "Empirical formula for static field ionization rates of atoms and molecules by lasers in the barrier-suppression regime," *J. Phys. B: At. Mol. Opt. Phys. 38 2593,* 2005.

[49] Zhang Q., Lan P., Lu P., "Empirical formula for over-barrier strong-field ionization," *Phys. Rev. A 90, 043410,* 2014.

[50] Remme S., Voitkiv A. B., Pretzler G., Müller C., "Phenomenological rate formulas for over-barrier ionization of hydrogen and helium atoms in strong constant electric fields.," *Journal of Physics B: Atomic, Molecular and Optical Physics, 58(19), 195602,* 2025.

[51] Bauer D., Mulser P., "Exact field ionization rates in the barrier-suppression regime from numerical time-dependent Schrödinger-equation calculations," *Phys. Rev. A 59(1) 569,* 1999.

[52] Kostyukov I. Y., Golovanov A. A., "Field ionization in short and extremely intense laser pulses," *Phys. Rev. A, 98(4), 043407,* 2018.

[53] Mironov A. A., Gelfer E. G., Tupitsyn I. I., Beck A., Jirka M., Klimo O., Meuren S., Oberreit G., Smorodnikova T., Taïeb R., et al., "Strong-field ionization in particle-in-cell simulations," *Physical Review E, 112(5), 055202,* 2025.

[54] Kolesik M., Moloney J. V. , "Nonlinear optical pulse propagation simulation: From Maxwell's to unidirectional equations," *Phys. Rev. E. 70, 036604 ,* 2004.

[55] Ptashnik I. V., Klimeshina T. E., Petrova T. M., Solodov A. A., Solodov A. M., "Water vapor continuum absorption in the 2.7 and 6.25 μm bands at decreased temperatures," *Atmos Ocean Opt, 29(3),* p. 211–215, 2016.